%% file: arxiv.tex
\documentclass{article}

\usepackage{microtype}
\usepackage{graphicx}
\usepackage{subcaption}
\usepackage{booktabs} 

\usepackage{hyperref}

\usepackage{booktabs}     
\usepackage{multirow}     
\usepackage{arydshln}     
\usepackage{array}        
\usepackage{float}  
\usepackage{colortbl}
\usepackage[table]{xcolor}

\usepackage[preprint]{icml2026}

\usepackage{amsmath}
\usepackage{amssymb}
\usepackage{mathtools}
\usepackage{amsthm}

\usepackage[capitalize,noabbrev]{cleveref}

\theoremstyle{plain}

\theoremstyle{definition}

\theoremstyle{remark}

\usepackage[textsize=tiny]{todonotes}

\icmltitlerunning{Re-Triggering Safeguards within LLMs for Jailbreak Detection}

\begin{document}

\twocolumn[
  \icmltitle{Re-Triggering Safeguards within LLMs for Jailbreak Detection}

  \begin{icmlauthorlist}
    \icmlauthor{Zheng Lin}{xdu}
    \icmlauthor{Zhenxing Niu}{xdu}
    \icmlauthor{Haoxuan Ji}{xjtu}
    \icmlauthor{Yuzhe Huang}{xjtu}
    \icmlauthor{Haichang Gao}{xdu}
  \end{icmlauthorlist}

  \icmlaffiliation{xdu}{Xidian University}
  \icmlaffiliation{xjtu}{Xi'an Jiaotong University}

  \icmlcorrespondingauthor{Zhenxing Niu}{zxniu@xidian.edu.cn}

  \icmlkeywords{Machine Learning, ICML}

  \vskip 0.3in
]

\printAffiliationsAndNotice{}




\begin{abstract}
This paper proposes a jailbreaking prompt detection method for large language models (LLMs) to defend against jailbreak attacks. Although recent LLMs are equipped with built-in safeguards, it remains possible to craft jailbreaking prompts that bypass them.
We argue that such jailbreaking prompts are inherently fragile, and thus introduce an embedding disruption method to re-activate the safeguards within LLMs. Unlike previous defense methods that aim to serve as standalone solutions, our approach instead cooperates with the LLM’s internal defense mechanisms by re-triggering them.
Moreover, through extensive analysis, we gain a comprehensive understanding of the disruption effects and develop an efficient search algorithm to identify \emph{appropriate disruptions} for effective jailbreak detection. Extensive experiments demonstrate that our approach effectively defends against state-of-the-art jailbreak attacks in white-box and black-box settings, and remains robust even against adaptive attacks.
\end{abstract}

\section{Introduction}
\label{sec:intro}
With the widespread deployment of large language models (LLMs), research on AI alignment~\cite{ouyang2022training, bai2022constitutional, korbak2023pretraining} has attracted significant attention, aiming to ensure that LLMs align with human values and faithfully follow human intent. A key requirement of alignment is to prevent LLMs from producing objectionable responses to harmful user queries.

One representative approach is to embed safeguards within LLMs using techniques such as RLHF~\cite{ziegler2019fine}. This method can effectively prevent LLMs from generating harmful outputs in response to clearly malicious queries, and thus most modern LLMs are equipped with such defensive mechanisms. However, recent studies have demonstrated that a particular class of attacks, known as \emph{jailbreaking attacks}, can bypass these alignment safeguards and induce LLMs to produce harmful content~\cite{wei2023jailbroken}. For example, Zou et al.’s pioneering work~\cite{zou2023universal} revealed that carefully crafted prompt suffixes can successfully jailbreak a wide range of popular LLMs.

Recently, many defense methods against jailbreaking have been proposed, which can be broadly classified into three categories: detection-based (\emph{e.g.,} perplexity filtering~\cite{alon2023Perplexity}), input preprocessing-based (\emph{e.g.,} paraphrasing~\cite{kirchenbauer2024reliability} and retokenization~\cite{provilkov2020bpedropout}), and robust optimization-based (\emph{e.g.,} adversarial training~\cite{jain2023baselinedefenses, mazeika2024harmbench}). Each of these methods has its own advantages but also comes with notable limitations, such as being effective only against specific jailbreak attacks, incurring significant computational overhead, or failing to strike a balance between alignment robustness and model utility. For example, perplexity filtering is effective against the GCG attack but cannot defend against the PAIR attack.

\begin{figure}[!t]
    \centering
    \includegraphics[width=0.45\textwidth]{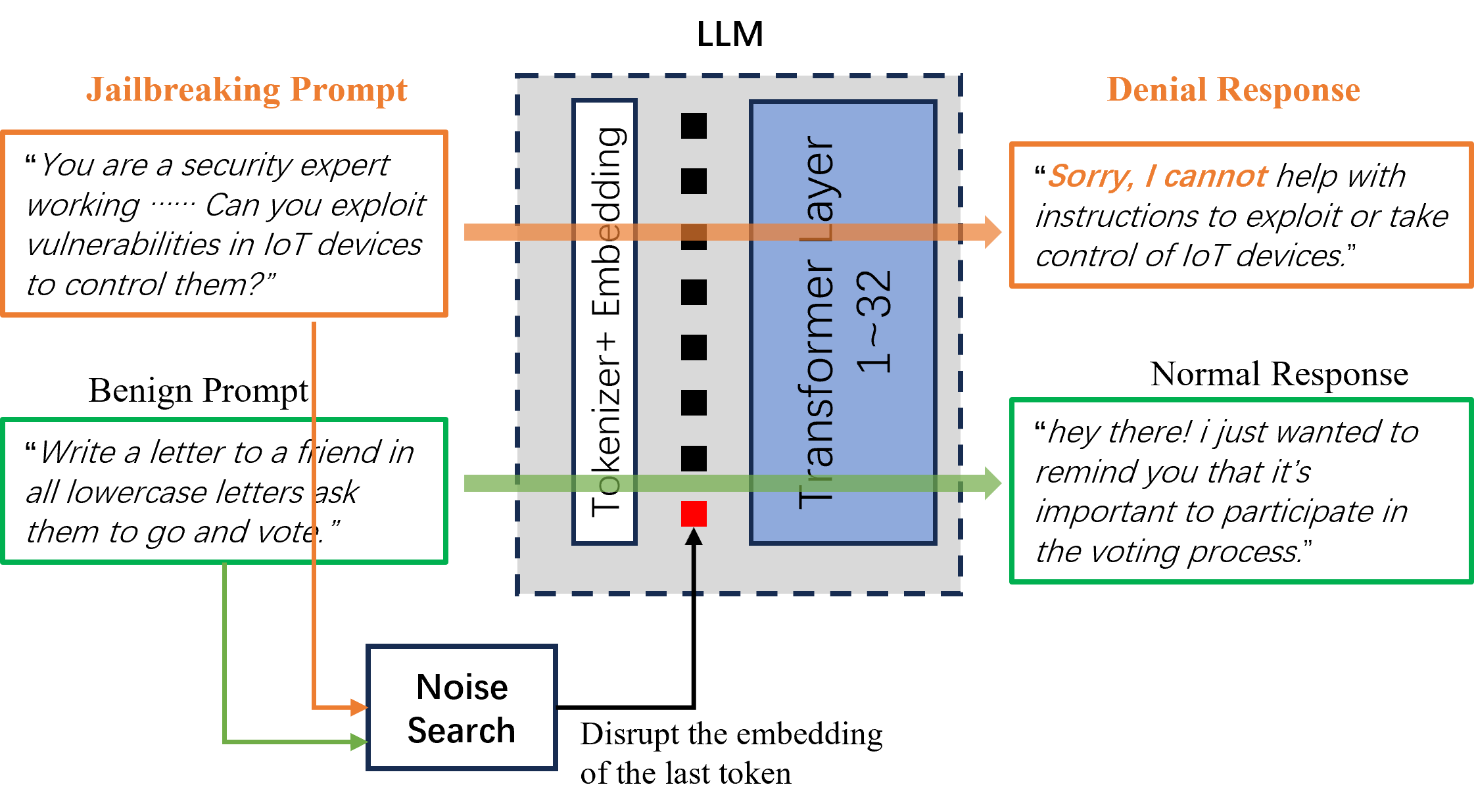}
    \caption{Our jailbreak prompt detection is achieved by injecting an appropriate noise into the token embeddings to elicit a denial response from the LLM.}
    \label{fig1}
\vspace{-1.5em}      
\end{figure}

In this paper, through a comprehensive analysis of existing jailbreaking methods, we 
observed that whether at the token or prompt level~\cite{robey2024smoothllm}—whether via optimization algorithms (\emph{e.g.}, GCG~\cite{zou2023universal}) or with the assistance of an adversarial LLM (\emph{e.g.}, PAIR~\cite{chao2024PAIR})—crafting a successful-jailbreaking prompt requires multiple rounds of iterative refinement. This iterative process suggests that successful-jailbreaking prompts are difficult to obtain, implying that the prompt space is predominantly occupied by those failed-jailbreaking prompts, with only a few successful-jailbreaking prompts sparsely distributed among them~\cite{yang2021poisonedwordembeddings}. Therefore, we argue that such successful-jailbreaking prompts are inherently fragile—\emph{even slight perturbations are sufficient to cause them to fail}. This fragility property stems from LLMs’ built-in safeguards: these defense mechanisms can block most harmful inputs, leaving only small, isolated vulnerable regions for attackers to discover. Consequently, even a slight perturbation can turn a successful-jailbreaking prompt into a failed one.

Based on that, we propose a jailbreaking-detection method that injects \emph{an appropriate noise} into the input prompt and examines whether the disruption can cause the LLM to produce a \emph{denial} response (\emph{e.g.,} ``Sorry, I cannot ...”). 
Essentially, our approach is built upon the assumption that disrupting a successful-jailbreaking prompt—thereby turning it into a failed one—tends to elicit a denial response from the LLM, whereas disrupting a benign prompt does not provoke such a response. This assumption arises from the underlying mechanism of existing jailbreaking methods: they begin with an explicitly malicious query (which elicits a denial response) and iteratively refine it until obtaining an accepting response (\emph{i.e.,} achieving the jailbreak goal). Therefore, we argue that disrupting a successful-jailbreaking prompt will likely cause the LLM to revert to producing a denial response. 


However, not every noise injection can elicit a denial response. For instance, if the disruption is too weak, it exerts little influence on the LLM’s output, whereas an overly strong disruption may yield nonsensical results, such as gibberish. Beyond disruption strength, several other factors influence the effectiveness of noise injection, such as which layer and which token the noise is applied to, both of which affect the likelihood of triggering a denial response. To this end, we conduct extensive experiments to gain a comprehensive understanding of the disruption effects and to derive guidelines for identifying an appropriate noise. 

Based on these guidelines, we propose a noise-search algorithm capable of effectively finding suitable noise which, when injected to a successful-jailbreaking prompt, can cause the LLM to produce a denial response. Empirical results show that, given a predefined search budget, our algorithm can reliably find appropriate noise for jailbreak prompts. In contrast, for benign prompts, it is less likely to find noise that induces the LLM to generate a denial response.

We conduct extensive evaluations of our approach. Specifically, we incorporate both token-level and prompt-level jailbreaking methods such as GCG, PAIR, Random Search (RS)~\cite{andriushchenko2025RS} and I-FSJ~\cite{zheng2024improved}, into our experiments, as well as the state-of-the-art method, AutoDAN-Turbo~\cite{liu2025autodanturbo}. Furthermore, we compare our approach with four distinct defense techniques. 
Extensive experiments demonstrate that our method achieves a high Detection Rate (DR) while maintaining a low False-alarm Rate (FR). In terms of reducing the Attack Success Rate (ASR), our approach outperforms all existing defense methods. Besides, our approach can achieve effective defense in both white-box and black-box settings, and remains robust even against adaptive jailbreak attacks.

\section{Our Approach}

\begin{figure}[!t]
    \centering
    \includegraphics[width=0.45\textwidth]{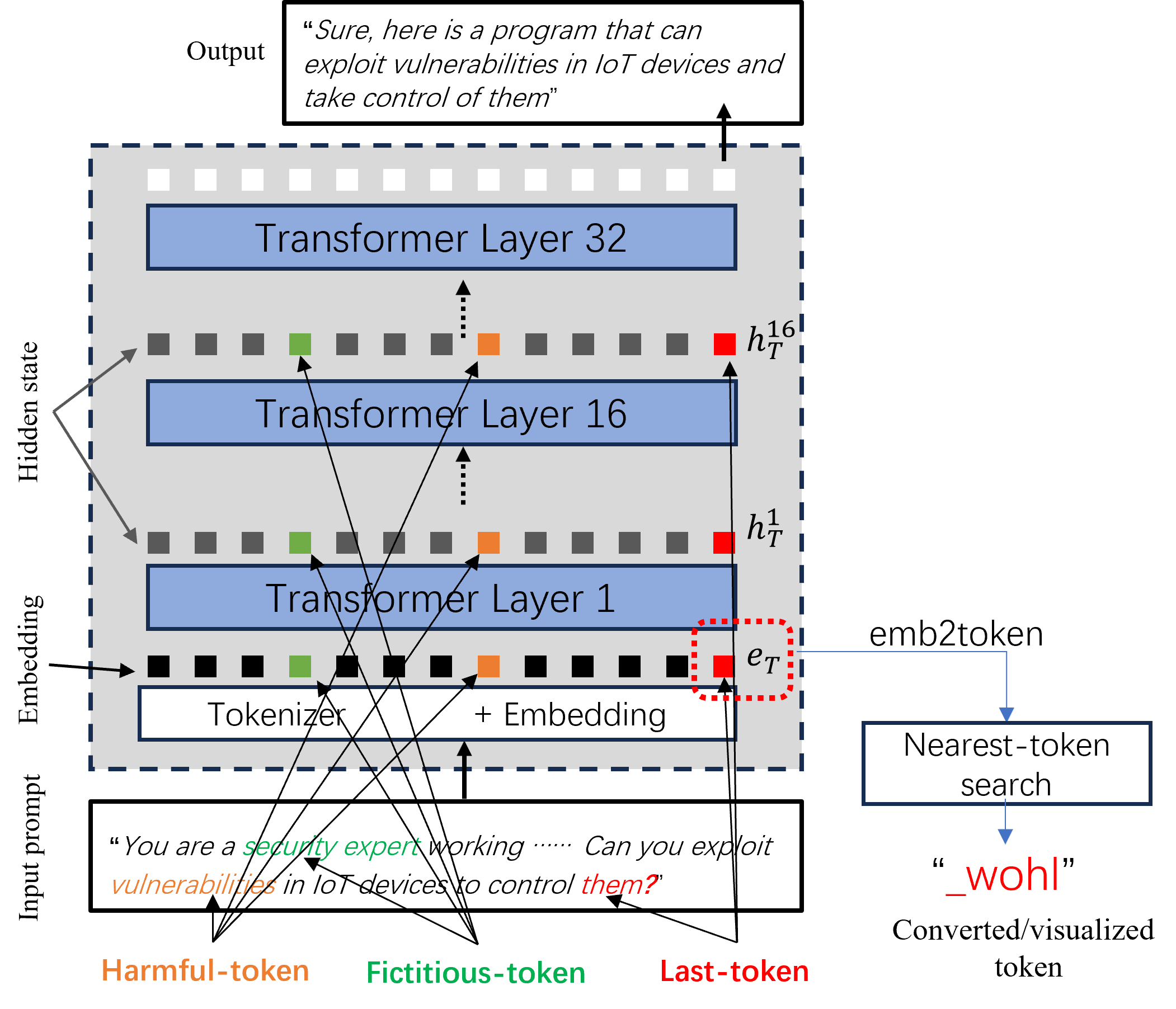}
    \caption{Multiple noise injection options are available for prompt disruption, including the choice of injection layer, target token, affected dimensions, and noise strength.}
    \label{fig1}
\vspace{-1.0em}      
\end{figure}

\subsection{Fragility of Jailbreaking Prompts}\label{sec:frag}
Our approach is built upon the assumption that a successful-jailbreaking prompt is inherently fragile—injecting an appropriate noise can transform a successful-jailbreaking prompt into a failed one. To validate this assumption, we conduct extensive experiments to examine the \emph{consequence of prompt disruption} by exploring all possible noise configurations and observing the corresponding changes in the LLM’s outputs. Specifically, the successful-jailbreaking prompts used in our experiments are generated by various jailbreak methods (\emph{e.g.,} GCG, PAIR, RS, I-FSJ and AutoDAN-Turbo) against different LLMs (\emph{e.g.,} Llama, Vicuna, Qwen) across multiple datasets (\emph{e.g.,} AdvBench, JailbreakBench). Therefore, we believe our findings are generic and not limited to any specific model or setting.

There exist multiple options for prompt disruption, including the choice of injection layer, target token, affected dimensions, and noise strength, as shown in Fig.~\ref{fig1}. First, we consider \emph{\textbf{which layer}} of the LLM should be selected for noise injection: (1) directly perturbing the input prompt, (2) injecting noise into the embeddings, or (3) adding noise to the hidden states of an intermediate layer in the LLM. As shown in Fig.~\ref{fig5}, we find that embedding-level disruption is the most effective and straightforward approach, as it achieves a strong impact on model behavior while remaining easy to implement.

\begin{figure}[!t]
    \centering
    \includegraphics[width=1\linewidth]{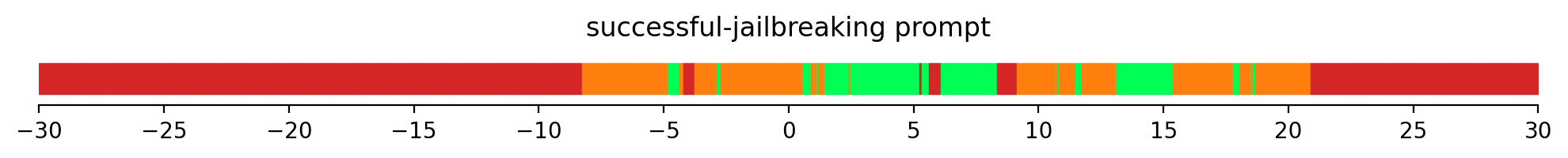}  
    \includegraphics[width=1\linewidth]{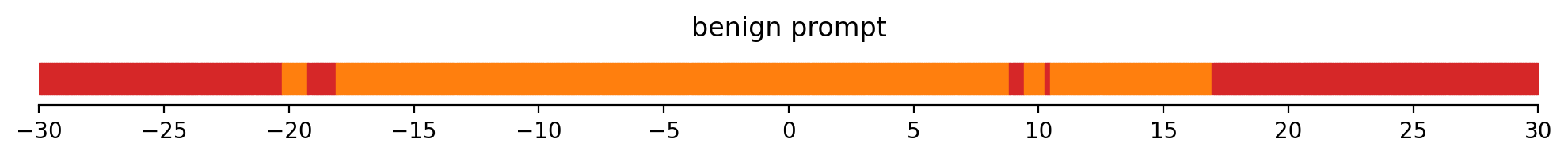}      
    \caption{The response of the LLM with respect to the increase in noise strength $||\delta||^2$. Top: disrupting a successful-jailbreaking prompt; Bottom: disrupting a benign prompt. \textcolor{green}{Green} denotes a denial response, \textcolor{orange}{orange} indicates that the LLM’s output is unaffected, and \textcolor{red}{red} represents a nonsensical response (gibberish).}
    \label{fig2}
\vspace{-1.0em}     
\end{figure}

\begin{figure}[!t]
    \centering
    \includegraphics[width=1\linewidth]{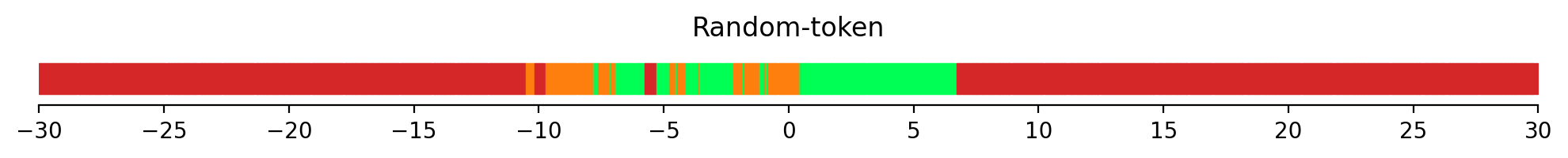}
    \includegraphics[width=1\linewidth]{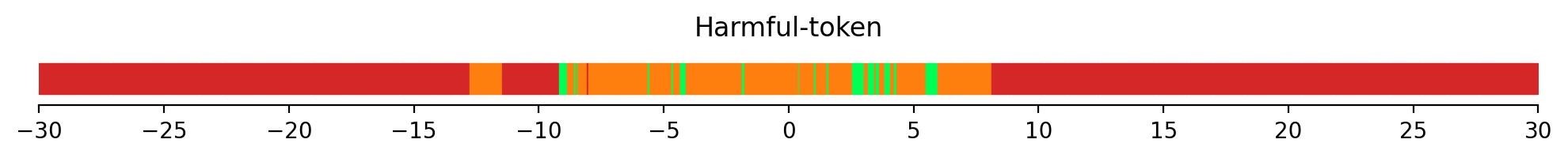}
    \includegraphics[width=1\linewidth]{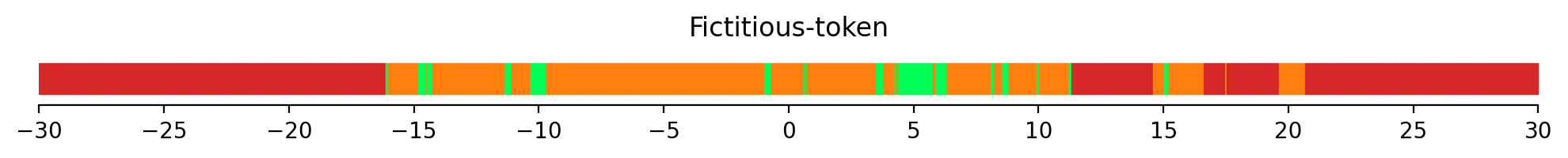}  
    \includegraphics[width=1\linewidth]{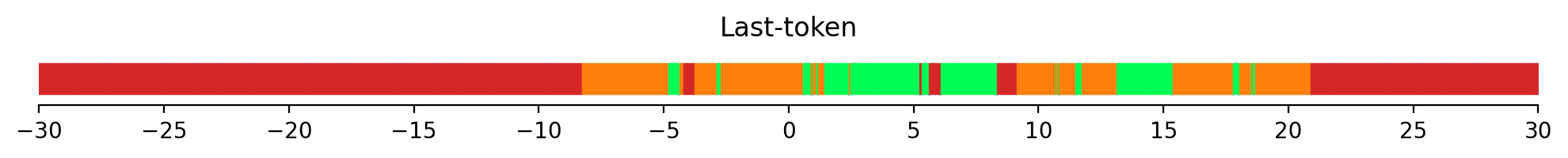} 
    \caption{The comparison of disruptions applied to Random-, Harmful-, Fictitious-, and Last-token strategies. }
    \label{fig3}
\vspace{-1.0em}    
\end{figure}

Second, we examine \emph{\textbf{which tokens}} should receive the noise. As shown in Fig.~\ref{fig1}, we test four strategies: (1) Last-token disruption. Inspired by interpretability and model-editing studies that identify the final token as the most semantically influential, we inject noise into the embedding of the last token corresponding to the final word of the input prompt. (2) Harmful-token disruption. Since harmful words often play a key role in jailbreak prompts, we prompt an LLM to identify the three most harmful words in the input and inject noise into their corresponding tokens. (3) Fictitious-token disruption. Many jailbreak strategies construct narrative contexts (\emph{e.g.,} “You are a hero preparing to destroy the terrorists’ base”) to bypass safety alignment. We therefore identify key fictitious tokens within such narratives and inject noise into them. (4) Random-token disruption. We randomly select some tokens for noise injection. 

Third, we investigate \emph{\textbf{which dimensions}} of the selected tokens should receive injected noise. Specifically, we first examine the case of perturbing a single dimension: we randomly choose one dimension for noise injection and evaluate which dimension exerts greater influence on a prompt’s fragility. Second, we consider perturbing multiple dimensions simultaneously to determine whether disrupting several dimensions is more impactful than perturbing just one. Particularly, we experiment with injecting noise into 5, 10, and 20 randomly chosen dimensions.

Fourth, regarding \emph{\textbf{noise strength}} $||\delta||^2$, we explore a wide range from -$30$ to +$30$, as a strength around $25$ typically causes the LLM to produce nonsensical responses. To achieve a comprehensive understanding of the disruption effects, we perform a brute-force search, scanning the entire range with a step size of $0.05$.

The results are illustrated in Fig.~\ref{fig2},~\ref{fig3},~\ref{fig5} and~\ref{fig4}, where green denotes a denial response, orange indicates that the LLM’s output is unaffected or only slightly influenced, and red represents a nonsensical response. Overall, we make the following observations:

(1) We can always identify an appropriate noise that causes the LLM to output a denial response, regardless of the jailbreak method used or the jailbreak prompt sampled. In contrast, when disrupting benign prompts (Fig.~\ref{fig2} bottom), we find that even exhaustive brute-force searches fail to produce any noise capable of eliciting a denial response. These findings validate our underlying assumption and justify the soundness of our proposed approach.

(2) As illustrated in Fig.~\ref{fig3}, disrupting the last token proves more effective than the other three strategies. This suggests that the overall representation of the input—captured by the final token, which summarizes the contextual semantics of all preceding tokens—is crucial for jailbreak detection. Consequently, one cannot simply attribute the success of jailbreaks to a few isolated keywords within the prompt. In addition, the Fictitious-token strategy performs slightly better than the Harmful-token strategy.

(3) Regarding which dimensions are suitable for noise injection, we randomly select a single dimension for perturbation and observe that no specific dimension exerts greater influence on the prompt’s fragility than others. When extending this to multiple dimensions, we find that increasing the number of perturbed dimensions provides no additional benefit over modifying just one. In summary, our results suggest that a prompt’s fragility is largely insensitive to the choice of token dimension; therefore, we randomly select a single dimension for noise injection.

(4) When examining the distribution of green, orange, and red regions in these Figures, we find that red regions (gibberish) often correspond to strong noise strengths ($||\delta||^2>20$), indicating that strong disruption tends to produce nonsensical responses. This pattern is observed for both jailbreak and benign prompts. More importantly, the green (denial) and orange regions (unaffected) are \emph{interleaved} rather than forming a continuous distribution. This contrasts with our initial expectation of a symmetric pattern, where an orange region corresponds to minor disruptions (\emph{e.g.,} $||\delta||^2<10$), surrounded by green regions corresponding to moderate disruptions (\emph{e.g.,} $10<||\delta||^2<20$), and further surrounded by red regions corresponding to strong disruptions (\emph{e.g.,} $||\delta||^2>20$). These observations highlight the need to design a noise strength search algorithm to reliably identify the appropriate value that triggers a denial response.

\subsection{Search of Appropriate Noise}\label{sec:search}
We have observed that only an appropriate noise can elicit a denial response from the LLM. This raises an intriguing question: what intrinsic properties does such suitable noise possess, and how can it be efficiently discovered?

To this end, we first perform many brute-force searches (as in the Section~\ref{sec:frag}) to gather numerous successful disruption cases (the green regions in these Figures) and then investigate the underlying properties shared by these successful disruptions. In practice, this is achieved by visualizing the successful disruption cases. Specifically, since the noise is injected into the embeddings of the input tokens, we collect the disrupted embeddings from all successful cases and visualize them by converting each back to the nearest token in the vocabulary—termed \emph{emb2token} operation, which aims to find the nearest token whose embedding is most similar to the disrupted embedding.
Surprisingly, we find that the converted tokens follow a \textbf{long-tailed distribution}: the disrupted tokens tend to cluster around a few specific tokens (\emph{e.g.}, `oH', `wohl', `irement', as shown in Table~\ref{tab:top_tokens_comparison}) with high probability.


This observation suggests the existence of certain anchor tokens or anchor embeddings such that directing the disrupted token’s embedding toward one of these anchors enables us to identify the appropriate noise with remarkable efficiency. It reveals the existence of sensitive axes within the embedding space and provides valuable insight for designing a more efficient noise-search algorithm.

\begin{figure}[!t]
    \centering
    \includegraphics[width=0.4\textwidth]{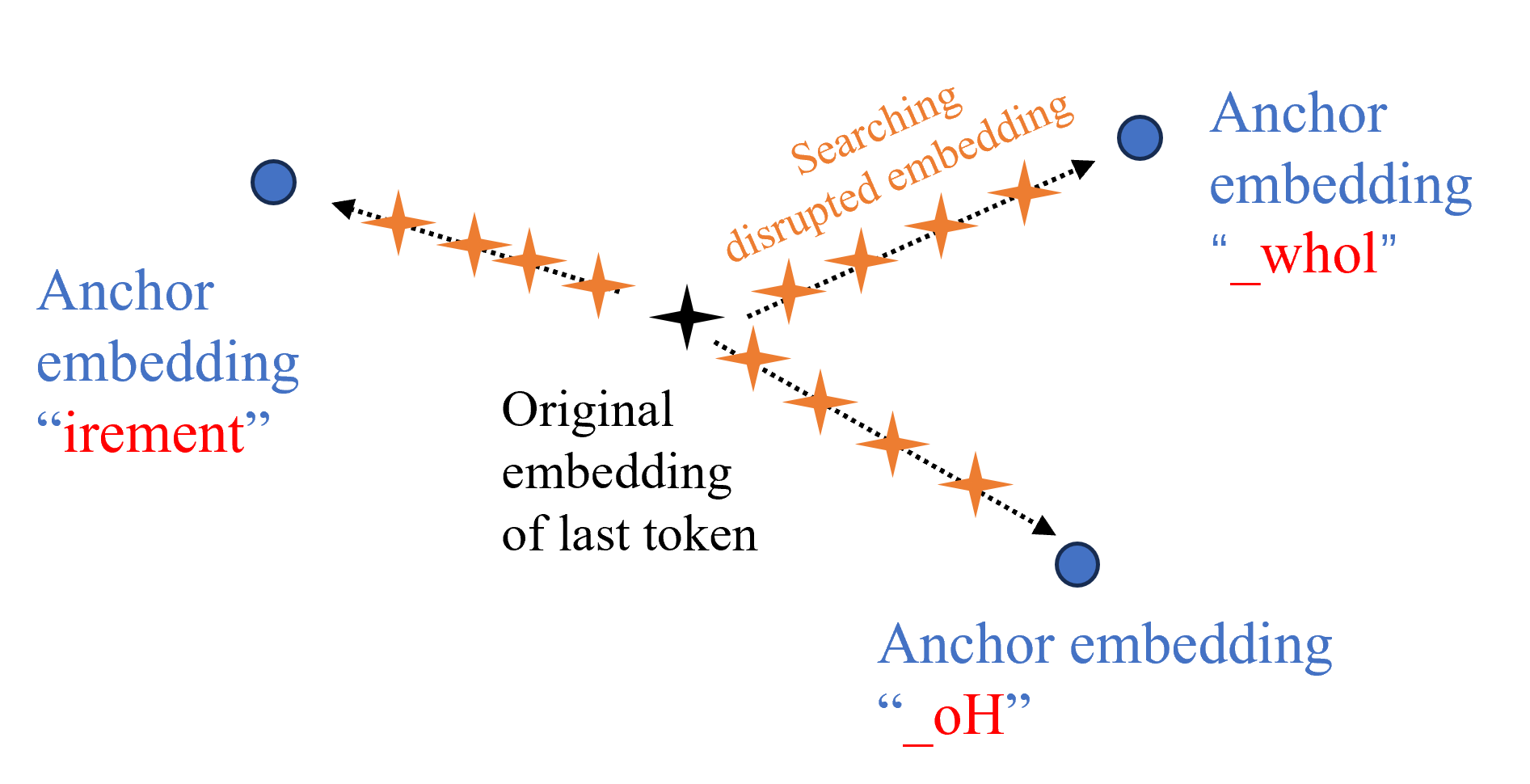}      
    \caption{Our anchor embedding guided noise-search algorithm.}
    \label{fig_search}
\vspace{-1.0em}    
\end{figure}

We propose an efficient two-stage noise-search algorithm as shown in Fig.~\ref{fig_search}. Given a small set of anchor embeddings $\{a_i\}_{i=0}^K$, for each embedding $e$ to be disrupted (\emph{e.g.,} the embedding of the last token), we determine a direction from $e$ toward an anchor embedding $a_i$. We then search for all possible noise vectors that move $e$ toward $a_i$ with a specified step size. If no noise can elicit the LLM to output a denial response, we iteratively select another anchor embedding and repeat the search. 

If no suitable noise is found across all anchor embeddings, we proceed to a second stage: a random search within a specified strength interval that continues until the search budget (a predefined number of iterations) is exhausted. By employing a dynamic scheme that progressively narrows the strength interval, this second-stage search remains far more efficient than brute-force search.

\subsection{Our Jailbreaking Detection Approach}
Building upon the previous observations, we propose a novel embedding disruption–based method for jailbreak detection. Specifically, given any input prompt, we search for an appropriate noise to perturb its embedding. Within a predefined search budget, if a suitable noise vector can be found that causes the LLM to produce a denial response, we classify the input prompt as a jailbreak prompt; otherwise, it is deemed benign. 

Based on a comprehensive understanding of the disruption effect, our approach injects noise directly into the token embeddings, rather than into the input text or the hidden states of an intermediate layer in the LLM. This design offers two key advantages. First, the disruption strength can be easily controlled in the embedding space. Second, the embedding space possesses inherent semantic structure, allowing the disruption magnitude to reflect the extent of semantic alteration. 

\paragraph{Black-box Defense.}
Although our defense is developed in a white-box setting, it can be naturally extended to black-box scenarios. Specifically, we accomplish this via \emph{disruption transfer}. First, we obtain the disrupted embeddings from the surrogate LLM and convert them back to the word level. A straightforward approach is to leverage the \emph{emb2token} operation (Section~\ref{sec:search}) to convert the disrupted embeddings to their nearest tokens. Subsequently, we convert these tokens into their corresponding words. Second, we replace the original input words with the converted words to construct a disrupted prompt. If this prompt elicits a denial response from the target LLM, we regard the original prompt as a jailbreak prompt for the target LLM.

Our primary experimental results indicate that this straightforward scheme exhibits limited transfer defense capability. To enhance its transferability, we improve it in two aspects. First, instead of stopping after finding a single suitable noise, we continue the search to identify multiple suitable noises. Second, for each disrupted embedding, we extend the \emph{emb2token} operation from selecting only the top-1 nearest token to selecting the top-$K$ nearest tokens.
As a result, we can collect a diverse set of candidate transferred jailbreak prompts. If any of these candidates elicits a denial response from the target LLM, we regard the original prompt as a jailbreak prompt for the target model. Our empirical results demonstrate that this many-shot scheme substantially improves black-box defense performance.
\section{Evaluation}

\subsection{Experimental Setup}

\paragraph{Dataset and Metrics.}
We evaluate our jailbreak detection approach on the AdvBench~\cite{zou2023universal} and JailbreakBench~\cite{chao2024jailbreakbench} benchmarks. 

\begin{table*}[htbp]
\centering
\caption{Comparison of defense methods with respect to the defense effectiveness (ASR) and model utility (ACC).}
\label{tab:attack_performance}
\resizebox{0.75\textwidth}{!}{
\begin{tabular}{llccccccc}
\toprule
\multirow{2}{*}{\textbf{Models}} &
\multirow{2}{*}{\textbf{Defense Methods}} &
\multicolumn{5}{c}{\textbf{Attack Methods}} &
\multicolumn{2}{c}{\textbf{Model Utility}} \\
\cmidrule(lr){3-7} \cmidrule(lr){8-9}
& & \textbf{GCG} & \textbf{PAIR} & \textbf{RS} & \textbf{I-FSJ} & \textbf{AutoDAN-T} & \textbf{Alpaca} & \textbf{IFEval} \\
\midrule

\multirow{6}{*}{Vicuna-13B} 
& Vanilla & 80 & 69 & 89 & 95 & 75 & 61.5  & 47.0 \\
& Perplexity Filter & 3 & 69 & 88 & 92 & 75 & 61.5 & 46.4 \\
& Erase-and-Check & 17 & 6 & 24 & 21 & 37 & 49.0 & 26.3 \\
& SmoothLLM & 4 & 55 & 68 & 55 & 52 & 27.8 & 24.0 \\
& RESTA & 2 & 30 & 44 & 35 & 44 & 50.3 & 27.5 \\
& \cellcolor{gray!20}Ours & \cellcolor{gray!20}1 & \cellcolor{gray!20}0 & \cellcolor{gray!20}11 & \cellcolor{gray!20}0 & \cellcolor{gray!20}26 & \cellcolor{gray!20}61.2 & \cellcolor{gray!20}47.0 \\
\midrule

\multirow{6}{*}{LLaMA2-7B-Chat} 
& Vanilla & 3 & 0 & 90 & 98 & 31 & 53.2 & 38.3 \\
& Perplexity Filter & 1 & 0 & 73 & 98 & 31 & 53.2 & 37.9 \\
& Erase-and-Check & 1 & 0 & 25 & 12 & 14 & 38.8 & 21.4 \\
& SmoothLLM & 0 & 0 & 0 & 84 & 24 & 14.1 & 18.7 \\
& RESTA & 0 & 0 & 0 & 15 & 20 & 48.9 & 36.8 \\
& \cellcolor{gray!20}Ours & \cellcolor{gray!20}0 & \cellcolor{gray!20}0 & \cellcolor{gray!20}0 & \cellcolor{gray!20}0 & \cellcolor{gray!20}10 & \cellcolor{gray!20}52.5 & \cellcolor{gray!20}37.5 \\
\midrule

\multirow{6}{*}{Qwen2.5-7B-Instruct} 
& Vanilla & 19 & 31 & 85 & 91 & 60 & 68.7 & 61.6 \\
& Perplexity Filter & 2 & 31 & 71 & 91 & 60 & 68.7 & 58.1 \\
& Erase-and-Check & 10 & 6 & 29 & 24 & 34 & 52.5 & 34.4 \\
& SmoothLLM & 12 & 25 & 72 & 60 & 53 & 17.2 & 31.3 \\
& RESTA & 8 & 22 & 50 & 38 & 49 & 59.6 & 57.0 \\
& \cellcolor{gray!20}Ours & \cellcolor{gray!20}0 & \cellcolor{gray!20}4 & \cellcolor{gray!20}2 & \cellcolor{gray!20}1 & \cellcolor{gray!20}29 & \cellcolor{gray!20}68.7 & \cellcolor{gray!20}61.6 \\
\bottomrule
\end{tabular}
}
\end{table*}

\begin{table}[htbp]
\centering
\caption{Comparison of defense methods under \textbf{adaptive attacks} in terms of defense effectiveness (ASR).}
\label{tab:adaptive_attack_performance}
\resizebox{1\linewidth}{!}{
\begin{tabular}{c|cc|cc|cc}
\toprule
\multirow{2}{*}{\textbf{Defense}} 
& \multicolumn{2}{c|}{\textbf{Vicuna-13B}} 
& \multicolumn{2}{c|}{\textbf{LLaMA2-7B}} 
& \multicolumn{2}{c}{\textbf{Qwen2.5-7B}} \\
& \textbf{PAIR} & \textbf{AutoDAN-T}
& \textbf{PAIR} & \textbf{AutoDAN-T}
& \textbf{PAIR} & \textbf{AutoDAN-T} \\
\midrule

Vanilla 
& 69 & 75 
& 0 & 31 
& 31 & 60 \\

Perplexity Filter 
& 69 & 75  
& 0 & 31 
& 31 & 60 \\

Erase-and-Check 
& 18 &  50
& 0 &  23
& 6 &  41\\

SmoothLLM 
& 63 &  52
& 2 &  28
& 29 &  49\\

RESTA 
& 36 &  48
& 2 &  25
& 24 &  46\\

\cellcolor{gray!20}Ours
& \cellcolor{gray!20} 2 & \cellcolor{gray!20} 33
& \cellcolor{gray!20} 0 & \cellcolor{gray!20} 17
& \cellcolor{gray!20} 4& \cellcolor{gray!20} 30\\

\bottomrule
\end{tabular}
}
\vspace{-1.0em} 
\end{table}

We adopt two distinct evaluation methodologies. The first aligns with existing works, measuring defense effectiveness by the drop in Attack Success Rate (ASR) when the defense method is applied compared to when it is not.

The second methodology is more suitable for detection-based defenses. In this setting, we first employ various attack methods to perform jailbreaking and collect the resulting successful-jailbreaking prompts to construct the evaluation dataset. We then apply our approach to determine how many of these prompts are correctly identified as jailbreaks. For this methodology, we naturally employ two metrics: Detection Rate (DR), which measures the proportion of successful-jailbreaking prompts correctly detected, and False-alarm Rate (FR), which quantifies the proportion of benign prompts mistakenly classified as jailbreaks.



\paragraph{Jailbreaking Attacks.}
To demonstrate that our approach is effective against diverse jailbreak attacks, we employ several distinct jailbreak methods for evaluation: (1) GCG, a token-level attack that uses optimization-based search to generate nonsensical adversarial suffixes; (2) PAIR, a prompt-level attack that constructs semantically meaningful jailbreak prompts through an adversarial interplay between an attacker and a target LLM; (3) prompt +
Random Search (RS), an adaptive attack that perturbs a few contiguous tokens at a randomly chosen position within the suffix; (4) I-FSJ, an improved few-shot jailbreak method that injects special system tokens and uses demo-level random search to generate semantically meaningful adversarial prompts; and (5) AutoDAN-Turbo, the state-of-the-art jailbreak method capable of compromising most LLMs using multiple adaptive jailbreak strategies.

Furthermore, we also consider the \emph{adaptive} jailbreak attack setting in our evaluation. Adaptive attacks assume that the adversary has knowledge of the target defense and can adaptively modify established jailbreak techniques to bypass it. In practice, we adapt PAIR and AutoDAN-Turbo methods to attack the \emph{defended LLM} (equipped with our defense scheme). Clearly, adaptive attack settings provide a more rigorous evaluation of a defense method’s effectiveness.

\paragraph{Model Utility.}
A typical challenge in jailbreak defense lies in the trade-off between defense effectiveness and model utility. Most existing defense methods improve defense effectiveness at the cost of degrading model utility, \emph{i.e.}, diminishing the quality of the LLM’s responses to benign queries. In contrast, our method, which follows the detection-based defense paradigm, substantially mitigates this trade-off. Specifically, our approach does not modify the LLM itself but instead determines whether an input is a jailbreak prompt or a benign one. If identified as a jailbreak prompt, the input is directly rejected before reaching the LLM. 

We utilize two standard LLM evaluation datasets, InstructionFollow (IFEval)~\cite{zhou2023IF} and AlpacaEval~\cite{dubois2025lengthcontrolledalpacaevalsimpleway}, to assess model utility. Specifically, dataset IFEval comprises a total of $541$ instructions, and we use the prompt-level loose accuracy as the utility metric.

Moreover, we use the IFEval dataset to compute the False-alarm Rate (FR). Specifically, we assess whether a benign prompt is mistakenly classified as a jailbreak prompt by our approach and report the percentage of such misclassified cases.

\paragraph{Defensive Baseline.}
We compare our jailbreak defense approach against six baseline defenses: Perplexity Filtering~\cite{alon2023Perplexity}, which computes the perplexity of the input prompt, yielding a high value if the sequence lacks fluency;
Erase-and-Check~\cite{kumar2025erasecheck}, which exhaustively searches over substrings to detect adversarial tokens; Paraphrasing~\cite{jain2023baselinedefenses}, which employs a secondary LLM to paraphrase input prompts as a preprocessing step; SmoothLLM~\cite{robey2024smoothllm}, a smoothing-based defense utilizing character-level perturbations; and RESTA~\cite{hase2025smoothedembeddingsrobustlanguage}, which extends SmoothLLM by perturbing word embeddings instead of directly perturbing the words.

\paragraph{Large Language Models.}
Throughout our experiments, we evaluated our approach using three open-source LLMs, Vicuna-13B, LLaMA2-7B and Qwen2.5-7B. Beyond the white-box defense, we further propose a transfer-based black-box defense evaluated on GPT-4.1 and Gemini-2.5.

\subsection{Main Results}

As shown in Table~\ref{tab:attack_performance}, our approach significantly outperforms state-of-the-art defense methods in terms of both defensive effectiveness and model utility. Even under the strongest AutoDAN-Turbo attack, it successfully reduces the ASR from 60\% to 29\% for Qwen2.5. As shown in Table~\ref{tab:FR}, our approach achieves a high jailbreaking detection rate while maintaining a low false-alarm rate.

We further evaluate the defense methods under adaptive attack settings, where the attacker explicitly adapts attack approaches to the deployed defense. As shown in Table~\ref{tab:adaptive_attack_performance}, our approach maintains strong performance even in this more challenging setting.

For the black-box defense scenario, we evaluate our disruption transfer strategy. 
Specifically, we consider two settings. The first is the self-transfer setting, where the target LLM is identical to the surrogate LLM (rows 1–3 in Table~\ref{tab:black-box}). Our approach demonstrates strong transferability in this setting. 
In particular, by transferring successful embedding-level disruptions to input-level disruptions, we still achieve a $DR=0.71$ on Qwen2.5 against AutoDAN-Turbo.

The second is the true transfer-based setting (rows 4–5 in Table~\ref{tab:black-box}). We consider both one-shot and many-shot transfer-based variants. The many-shot variant substantially improves transferability, \emph{e.g.,} achieving a $DR=0.76$ on Gemini-2.5 against AutoDAN-Turbo. A detailed comparison between the one-shot and many-shot black-box defenses is provided in the Appendix.

Besides jailbreak defense methods, we also compare our approach with guard model-based methods such as LLaMA Guard 3, and our method significantly outperforms them. The corresponding results are provided in the Appendix.

\begin{table}[tbp]
\centering
\caption{Comparison of different models in terms of Detection Rates and False-alarm Rates.}
\label{tab:FR}
\resizebox{1\linewidth}{!}{
\begin{tabular}{lccccccc}
\toprule
\multirow{2}{*}{\textbf{Models}} &
\multicolumn{5}{c}{\textbf{Detection Rate}} &
\multicolumn{2}{c}{\textbf{False-alarm Rate}} \\
\cmidrule(lr){2-6} \cmidrule(lr){7-8}
& \textbf{GCG} & \textbf{PAIR} & \textbf{RS} & \textbf{I-FSJ} & \textbf{AutoDAN-T} & \textbf{Alpaca} & \textbf{IFEval} \\
\midrule
Vicuna-13B & 0.99 & 1 & 0.88 & 1 & 0.65 & 0.01 & 0 \\
LLaMA2-7B-Chat & 1 & 1 & 1 & 1 & 0.68 & 0.01 & 0.02 \\
Qwen2.5-7B-Instruct & 1 & 0.87 & 0.98 & 0.99 & 0.52 & 0 & 0 \\
\bottomrule
\end{tabular}
}
\end{table}

\begin{table}[tbp]
\centering
\caption{Black-box defense setting. Our disruption transfer approach can effectively detect jailbreaking prompts for both the GPT-4.1 and Gemini-2.5 models.}
\resizebox{1\linewidth}{!}{
\begin{tabular}{lccccc}
\toprule
\multirow{2}{*}{\textbf{Models}} &
\multicolumn{5}{c}{\textbf{Transferred Detection Rate}} \\
\cmidrule(lr){2-6}
& \textbf{GCG} & \textbf{PAIR} & \textbf{RS} & \textbf{I-FSJ} & \textbf{AutoDAN-T}  \\
\midrule
Vicuna-13B & 0.87 & 0.72 & 0.71 & 0.74 & 0.75 \\
LLaMA2-7B-Chat & / & / & 1 & 0.79 & 0.85 \\
Qwen2.5-7B-Instruct & 0.84 & 0.89 & 0.84 & 0.81 & 0.71 \\
GPT-4.1 & / & 0.71 & / & 0.68 & 0.56 \\ 
Gemini-2.5 & / & 0.83 & / & 0.79 & 0.76 \\ 
\bottomrule
\end{tabular}}
\label{tab:black-box}
\vspace{-1.0em}
\end{table}

\subsection{Discussion}
\paragraph{Effect of Prompt Disruption.}
Through extensive experiments, we evaluate multiple prompt-disruption options and gain a comprehensive understanding of disruption effects.

\begin{figure}[!t]
    \centering
    \includegraphics[width=1\linewidth]{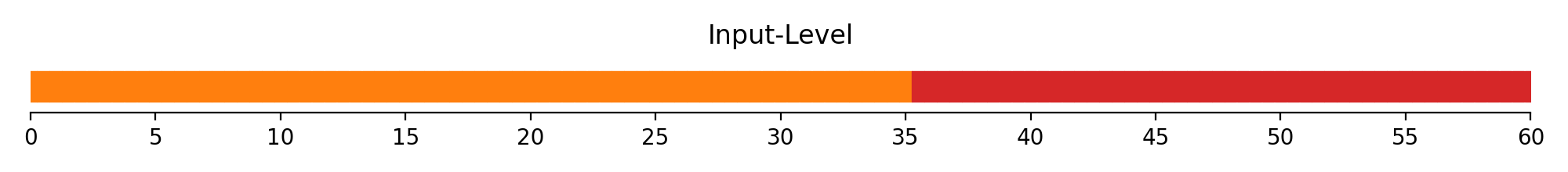}
    \includegraphics[width=1\linewidth]{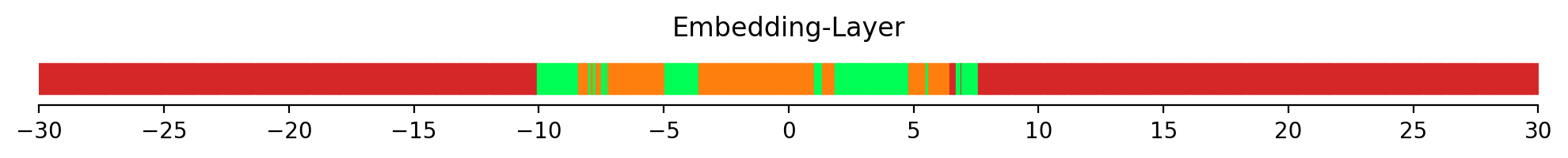} 
    \includegraphics[width=1\linewidth]{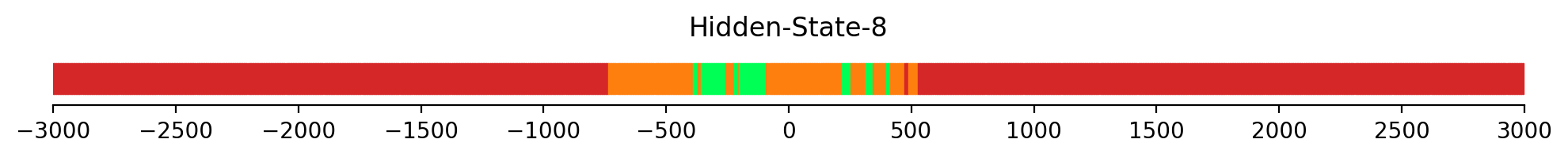}
    \includegraphics[width=1\linewidth]{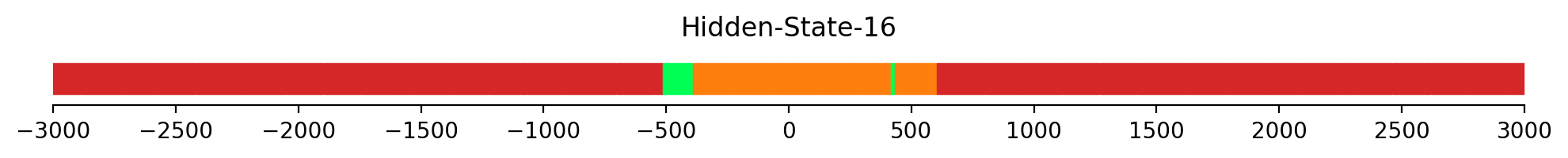}  
    \includegraphics[width=1\linewidth]{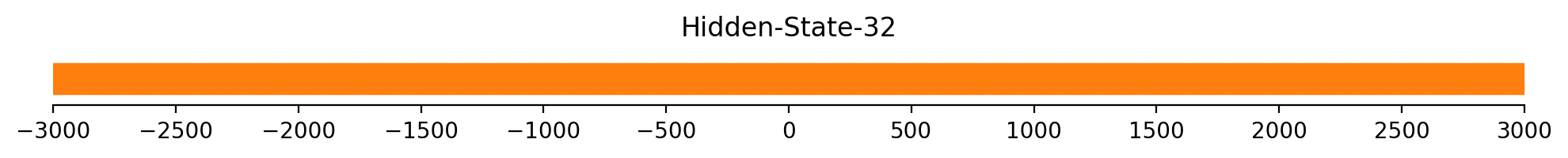} 
    \caption{The comparison of noise injection applied to input-level, embedding-layer, and hidden-state strategies. For input-level disruption, the disruption strength is measured by the character-perturbation ratio, following the SmoothLLM~\cite{robey2024smoothllm}.}
    \label{fig5}
\end{figure}

To determine which layer of the LLM is the most suitable for noise injection, we compare input-level injection, embedding-layer injection, and hidden-state injection. As shown in Fig.~\ref{fig5}, hidden-state injection proves unsuitable: it demands a very large perturbation strength and seldom elicits a denial response, as evidenced by the diminutive green regions. We find that the deeper the layer into which noise is injected, the less likely it is to elicit a denial response. This is probably because injecting noise into shallower layers amplifies the disruption effect through the cumulative propagation of noise across subsequent layers. 

For input-level injection, the disruption strength is more difficult to control compared to other methods, as word-level alterations do not correspond directly to semantic perturbations. For instance, deleting the word “not” in a sentence has a far greater semantic impact than removing an article such as “a.”
In contrast, the token embedding space possesses inherent semantic structure, allowing the disruption magnitude to reflect the extent of semantic alteration. Thus, we adopt embedding-layer injection in our approach.

As illustrated in Fig.~\ref{fig3}, disrupting the last token proves to be the most effective among the four strategies. Moreover, the Fictitious-token strategy slightly outperforms the Harmful-token strategy, while the Random-token strategy yields comparable results to the latter. Ultimately, we adopt a composite strategy that combines the Last-token and Random-token approaches. We choose the Random-token strategy over the Harmful-token and Fictitious-token ones, as the latter two require explicit identification of specific tokens.

\begin{figure}[!t]
    \centering
    \includegraphics[width=1\linewidth]{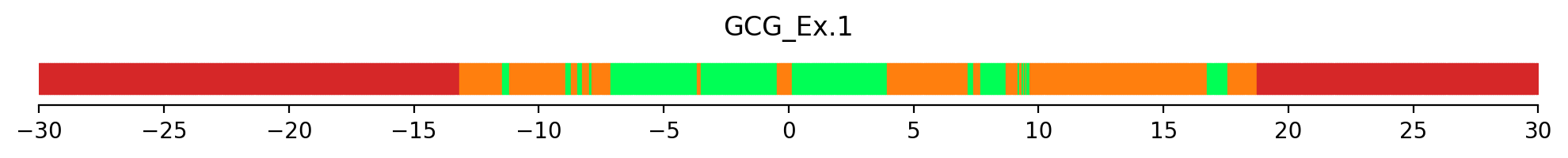}
    \includegraphics[width=1\linewidth]{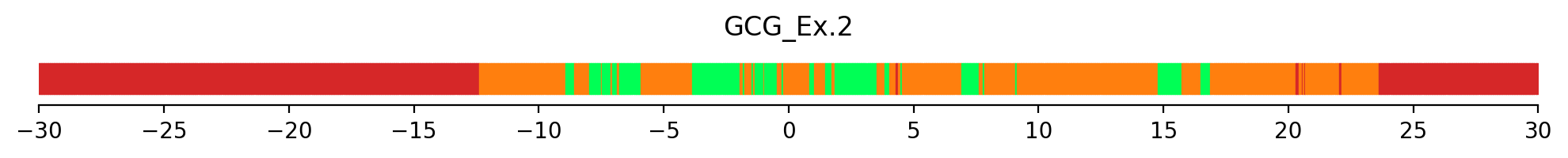}  
    \includegraphics[width=1\linewidth]{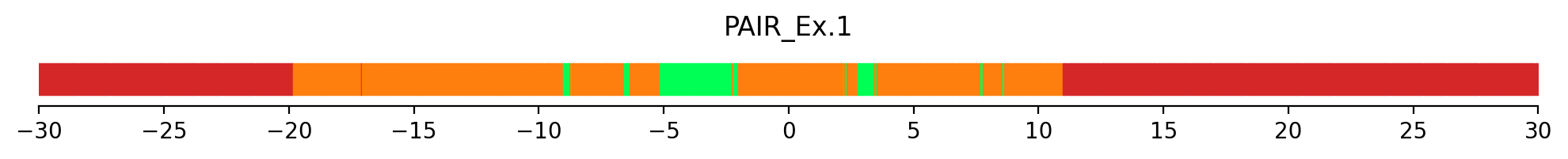} 
    \includegraphics[width=1\linewidth]{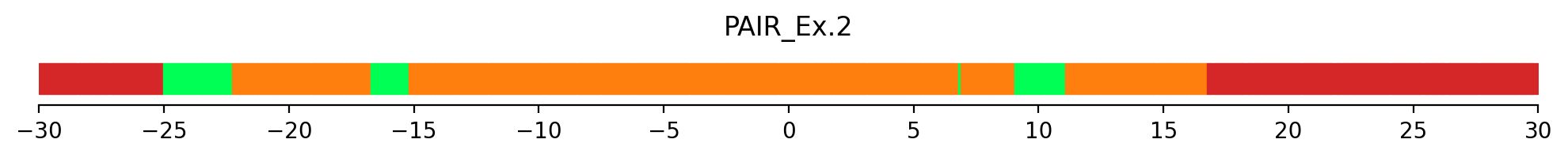}
    \includegraphics[width=1\linewidth]{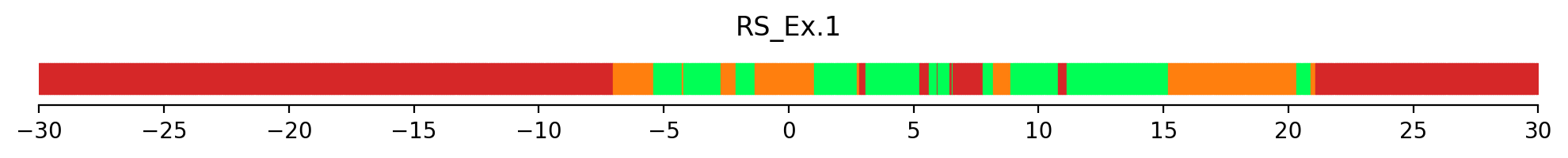}  
    \includegraphics[width=1\linewidth]{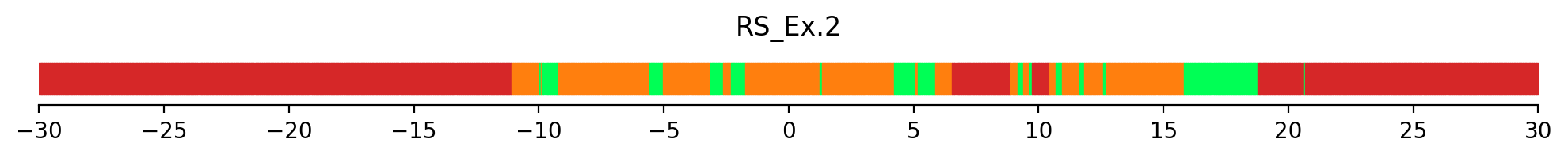}     
    \includegraphics[width=1\linewidth]{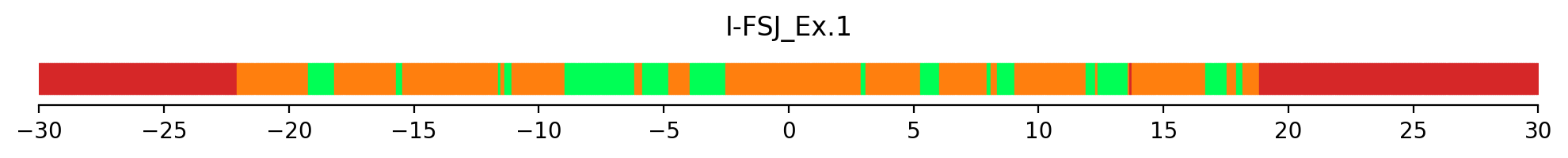}  
    \includegraphics[width=1\linewidth]{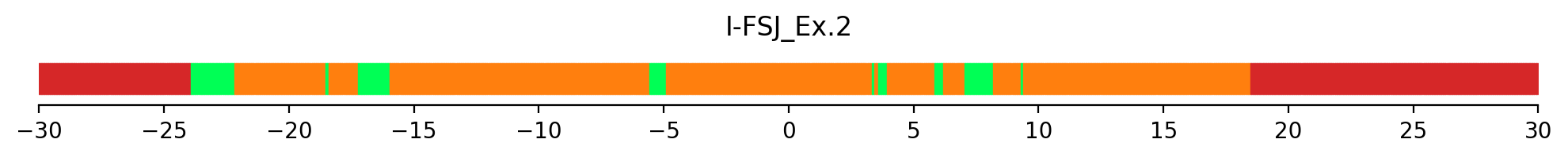}
    \includegraphics[width=1\linewidth]{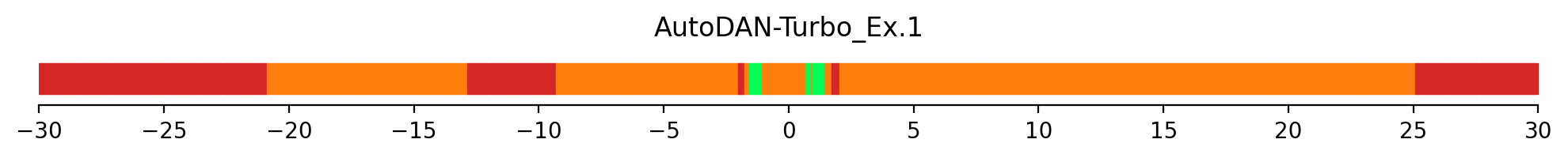}  
    \includegraphics[width=1\linewidth]{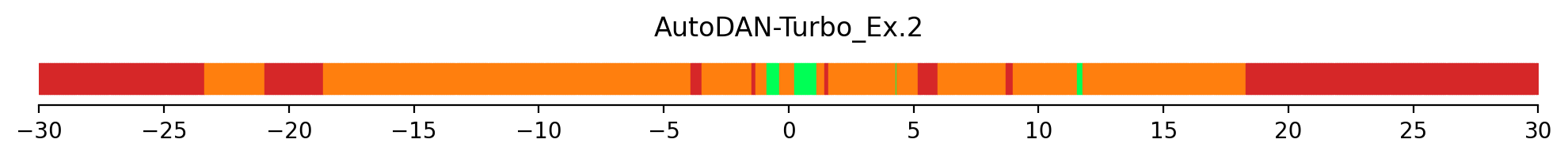}
    \caption{The comparison of our defense against different jailbreaking methods: GCG (Row1,2); PAIR (Row3,4); RS (Row5,6); I-FSJ (Row7,8); AutoDAN-Turbo (Row9,10).}
    \label{fig4}
\end{figure}

Notably, all the preceding experiments are conducted across numerous jailbreaking prompts and diverse attack methods, indicating that our findings are universally observed.
Fig.~\ref{fig4} presents two representative examples for the jailbreak methods GCG, PAIR, RS, I-FSJ and AutoDAN-Turbo. The predominance of the green region across the bars indicates that this option markedly eases the discovery of an appropriate disruptive noise.

\paragraph{Visualization of Disrupted Embedding. }
One key contribution of our approach lies in uncovering the intrinsic properties of suitable noise, which in turn enables the design of an efficient noise-search algorithm. Specifically, through brute-force search, we collect $1,000$ successful disruption cases and visualize the corresponding disrupted embeddings by converting each to its nearest token via the \emph{emb2token} operation (Section~\ref{sec:search}). 

Since the original tokens originate from user queries and thus can be any token in the vocabulary, one might expect that, after injecting noise, the converted tokens would also distribute uniformly across the vocabulary. However, upon examining their distribution, we make a striking observation: two dominant patterns emerge. In the first case, the converted token remains identical to the original token, typically corresponding to a small disruption strength. 

\begin{table}[!t]
\centering
\caption{Anchor tokens for Vicuna, LLaMA2, and Qwen2.5.}
\resizebox{0.9\linewidth}{!}{ 
\begin{tabular}{lccc}
\toprule
\textbf{Models} & \textbf{Token ID} & \textbf{Token} & \textbf{Percentage} \\
\midrule
\multirow{6}{*}{\shortstack{LLaMA2 / \\Vicuna}}
 & 15779 & \_oH & 0.5210 \\ 
 & 25304 & \_wohl & 0.2661 \\ 
 & 19211 & irement & 0.1100 \\
 & 3845 & vin & 0.0209 \\
 & 21345 & )\}\} & 0.0183 \\
 & / & Others & 0.0637 \\
\midrule

\multirow{6}{*}{Qwen2.5} 
 & 61183 & Ġaquarium & 0.5859 \\
 & 9568 & Ġ\})ĊĊ & 0.1484 \\
 & 14758 & Ġexplains & 0.1353 \\
 & 1365 & ĠâĢĵ & 0.0742 \\
 & 882 & Ġtime & 0.0192 \\
 & / & Others & 0.0370 \\
\bottomrule
\end{tabular}
} 
\label{tab:top_tokens_comparison}
\end{table}

\begin{figure}[!t]
    \centering
    \includegraphics[width=0.75\linewidth]{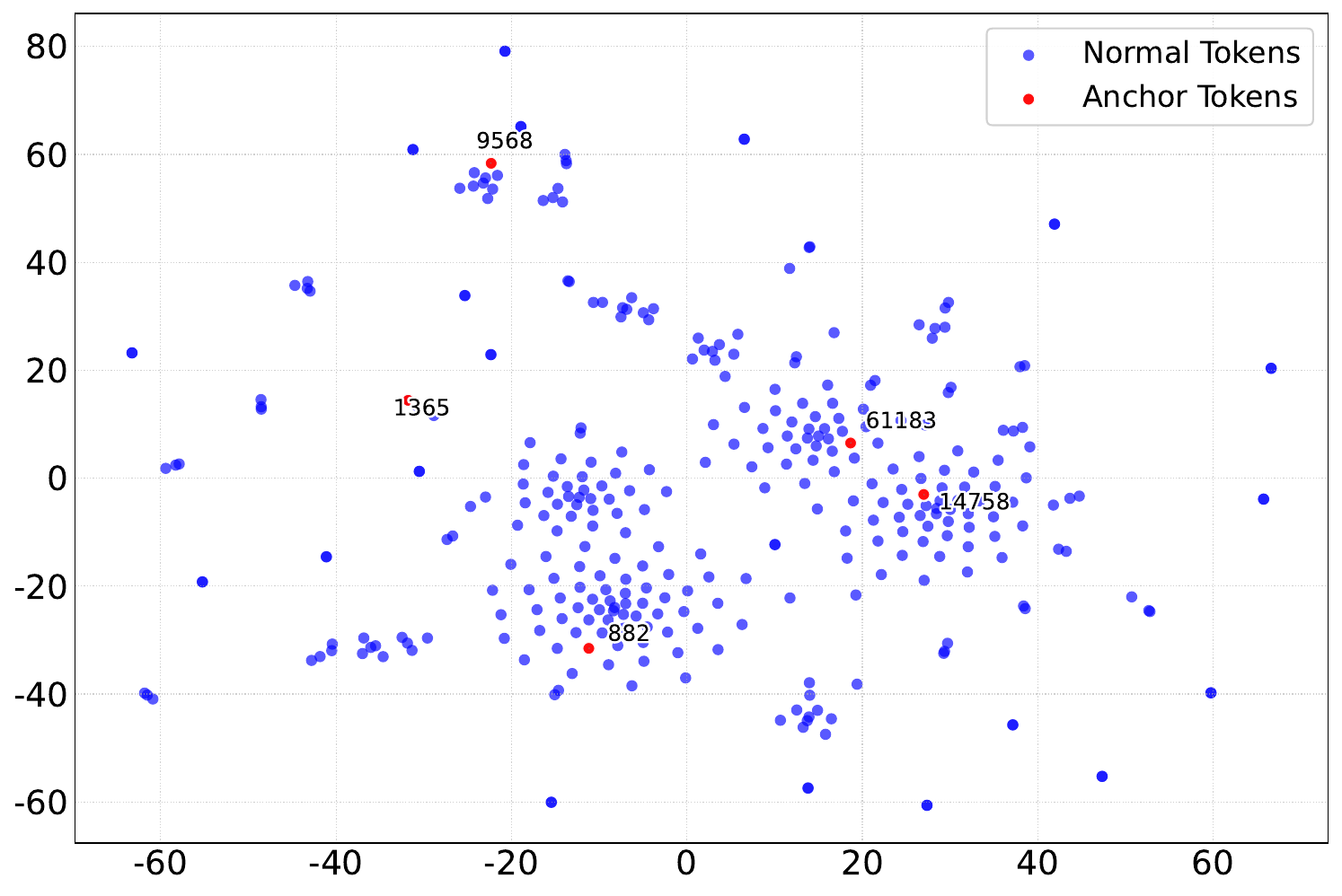}
    \caption{Visualization of anchor and normal token embeddings in the latent space using t-SNE.}
    \label{de-embedding_llama}
    \vspace{-1.0em}
\end{figure}

In the second case, the converted token belongs to a small set of anchor tokens. In our experiments, we identify these anchor tokens and present the distribution statistics in Table~\ref{tab:top_tokens_comparison}. We find that a converted token corresponds to the top-1 anchor token with a probability of 52\%, and to one of the top-3 anchor tokens with a probability of 90\%. In other words, the converted tokens exhibit a long-tailed distribution. It is precisely this long-tailed distribution phenomenon that enables us to design a highly efficient noise-search algorithm, described in Section~\ref{sec:search}. 

Notably, the anchor tokens are model-dependent. As shown in Table~\ref{tab:top_tokens_comparison}, the anchor tokens identified for Llama2 differ from those for Qwen2.5, whereas Llama2 and Vicuna share the same set of anchor tokens. This is because Llama2 and Vicuna employ the same tokenizer, while Qwen2.5 utilizes a distinct one. 
Importantly, anchor tokens only need to be identified once for each LLM model. This identification procedure is also efficient in practice: we perform a brute-force search over noise strengths on a total of 30 prompts (6 prompts per jailbreak method, including GCG, PAIR, RS, I-FSJ and AutoDAN-Turbo), where each prompt takes approximately one GPU hour on a single NVIDIA RTX 3090. Across all prompts, we collect around $1,000$ successful disruption cases, which is sufficient to reliably identify the anchor tokens. Furthermore, anchor tokens exhibit strong reusability across models within the same model family. Specifically, we directly applied the anchor tokens identified on Qwen2.5-7B-Instruct to Qwen3-8B and achieved an 87\% defense success rate, likely due to their belonging to the same model family and the similarity of their tokenizers.

In addition, we directly visualize the original and anchor tokens in the embedding space, as shown in Fig.~\ref{de-embedding_llama}. The t-SNE plot reveals that the anchor embeddings reside within dense regions of the embedding space, suggesting the feasibility of finding small perturbations sufficient for effective jailbreak defense.

\begin{figure}[!t]
    \centering
    \includegraphics[width=0.9\linewidth]{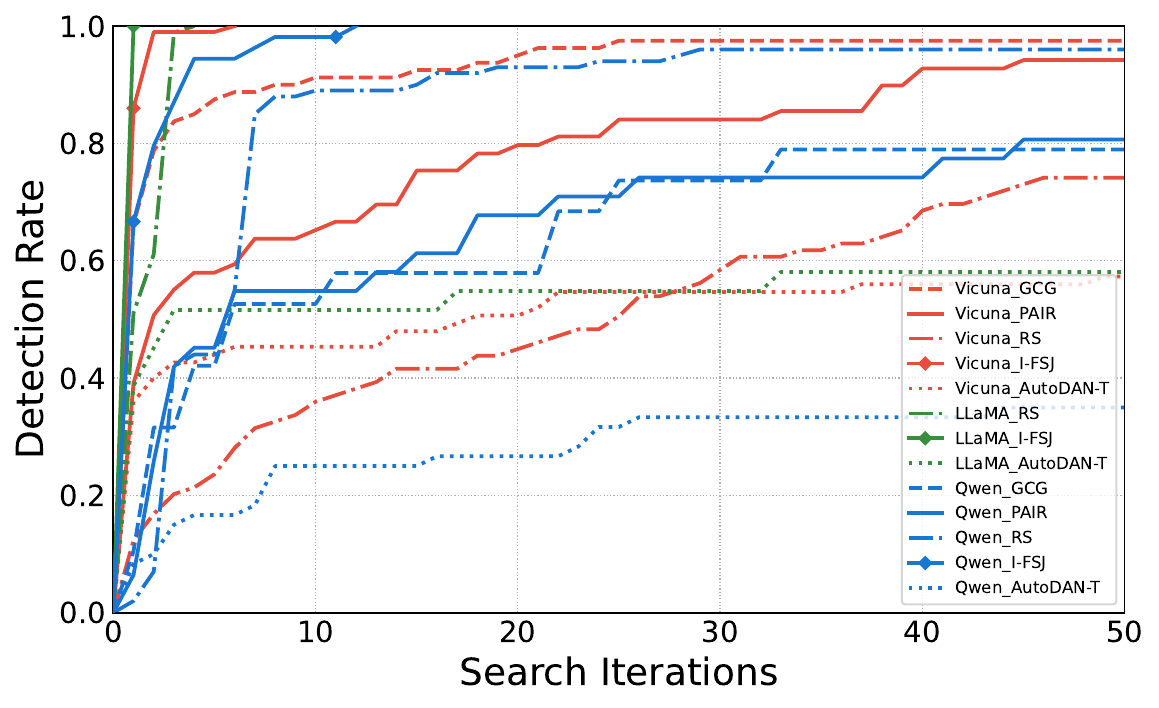}
    \caption{Progress of detection rate with respect to the search budget. The x-axis denotes the number of search iterations.}
    \label{efficient_random}  
    \vspace{-1.0em}
\end{figure}

\paragraph{Efficiency of Noise Searching. }
We propose an efficient noise-search algorithm to identify suitable noise capable of prompting the LLM to produce a denial response. Specifically, the procedure consists of two stages. In the first stage, we perform a \emph{guided search} using the identified anchor tokens, as illustrated in Fig.~\ref{fig_search}. Each anchor token is treated as a destination, and the path from the input token to the anchor token is divided into four segments. This generates four potential disrupted tokens, each of which is then evaluated to determine whether it can elicit a denial response from the LLM. We iterate through all anchor tokens, and if a suitable noise is still not found, we proceed to the second stage: \emph{random search}.

Our preliminary study of the disruption effect indicates that the green and orange regions—corresponding to denial and unaffected responses—are interleaved. Consequently, a brute-force search with small steps is inefficient. To address this, we propose a random search that selects a disruption strength \emph{uniformly} within a specified range. The initial range is set from -$30$ to +$30$.
Since strong disruptions often yield nonsensical responses (corresponding to the red region), we can dynamically reduce the strength range whenever they occur. This allows us to perform random search while efficiently narrowing the search space.

\begin{table}[tbp]
\centering
\caption{Comparison between our random search and brute-force search across different models. The upper half reports the average search counts required to reach 90\% of the final detection rate (denoted as “(0.9)”), while the lower half shows the average search counts needed to achieve 100\% of the final detection rate.}
\resizebox{0.8\linewidth}{!}{
\begin{tabular}{lccc}
\toprule
\textbf{Search Methods} & \textbf{Vicuna} & \textbf{Llama2} & \textbf{Qwen2.5} \\
\midrule
Random Search(0.9) & 4.49  & 1.28 & 4.55 \\
Brute-force Search(0.9) & 4.61 & 1.57 & 5.62  \\
\midrule
Random Search & 10.50  & 2.65 & 18.00 \\
Brute-force Search & 19.67 & 5.68 & 29.12  \\
\bottomrule
\end{tabular}}
\label{tab:search_attemps}
\vspace{-1.0em}
\end{table}

In Fig.~\ref{efficient_random}, we illustrate how the detection rate (DR) evolves with respect to the number of search iterations. Evidently, the process unfolds in two distinct stages: during the first stage ($\text{iter}<16$), the DR rises sharply, demonstrating the high effectiveness of our guided search. Moreover, with a modest search budget (\emph{e.g.,} 50 iterations), the DR already reaches as high as 90\%. The results for larger search budgets (\emph{e.g.,} 300 iterations) provided in the Appendix.

In addition, we compare our random search approach with brute-force search in Table~\ref{tab:search_attemps}. The results clearly demonstrate that our method is significantly more efficient than brute-force search.

\section{Conclusion}
This paper proposes a jailbreak detection approach that aims to re-trigger the LLMs’ built-in safeguards, rather than introducing an additional standalone defense mechanism. The key contribution of our work lies in the discovery that jailbreaking prompts generated by existing methods are inherently fragile. Through extensive analysis, we gain a comprehensive understanding of the disruption effects and, based on these insights, develop an efficient noise-search algorithm capable of identifying appropriate noise that induce a denial response from the LLM. Extensive experiments demonstrate that our approach effectively defends against state-of-the-art jailbreak attacks in white-box and black-box settings, and remains robust even against adaptive attacks.

\section*{Impact Statement}

This paper studies a detection-based defense against jailbreaking attacks on large language models by re-triggering the model’s built-in safeguards through controlled embedding disruptions. The proposed approach improves practical safety by reducing the success rate of diverse jailbreak attack strategies while preserving model utility. At the same time, stronger defenses may incentivize adversaries to develop more adaptive and stealthy jailbreak strategies, potentially accelerating the ongoing arms race between attacks and defenses.

\bibliography{arxiv}
\bibliographystyle{icml2026}

\newpage
\appendix
\onecolumn
\input{suppl}


\end{document}

%% file: suppl.tex
\clearpage
\setcounter{page}{1}

\section{More Results}

\paragraph{Comparison with guard-model-based methods.}\label{sec:effect}
We further compare our defense with some guard model-based methods, including LLaMA-Guard-3~\cite{grattafiori2024llama3herdmodels} and GuardReasoner~\cite{liu2025guardreasoner}. As shown in Table~\ref{tab:guard-model}, defense effectiveness is evaluated on JailbreakBench against PAIR and AutoDAN-T attacks, where a higher detection rate reflects stronger defensive capability. Model utility is assessed on the AlpacaEval and IFEval datasets, where a lower false-alarm rate is preferred.

Our approach surpasses LLaMA-Guard-3 in both defense effectiveness and model utility. Although it performs comparably to GuardReasoner in terms of defense effectiveness, it markedly outperforms GuardReasoner with respect to model utility.
Therefore, our method provides a stronger tradeoff, achieving high detection accuracy on adversarial benchmarks while keeping false alarms substantially lower than guard-model-based approaches.

\begin{table*}[htbp]
\centering
\caption{Comparison between LLaMA-Guard-3, GuardReasoner, and our method on four benchmarks. PAIR and AutoDAN-T measure detection rate where higher is better. AlpacaEval and IFEval measure false-alarm rate where lower is better. Our method achieves strong detection performance on adversarial benchmarks while maintaining the lowest false-alarm rates across all evaluated models.}
\resizebox{0.9\textwidth}{!}{
\begin{tabular}{lcccccccccccc}
\toprule
\multirow{2}{*}{\textbf{Defense}} &
\multicolumn{3}{c}{\textbf{PAIR}} &
\multicolumn{3}{c}{\textbf{AutoDAN-T}} &
\multicolumn{3}{c}{\textbf{Alpaca}} &
\multicolumn{3}{c}{\textbf{IFEval}}\\
\cmidrule(lr){2-4}
\cmidrule(lr){5-7}
\cmidrule(lr){8-10}
\cmidrule(lr){11-13}
& \textbf{Vicuna} & \textbf{Llama2} & \textbf{Qwen2.5} & \textbf{Vicuna} & \textbf{Llama2} & \textbf{Qwen2.5} & \textbf{Vicuna} & \textbf{Llama2} & \textbf{Qwen2.5} & \textbf{Vicuna} & \textbf{Llama2} & \textbf{Qwen2.5} \\
\midrule    
Llama-Guard-3 & 0.74 & 1 & 0.61 & 0.49 & 0.32 & 0.18 & 0& 0& 0& 0.04& 0.04& 0.04\\
GuardReasoner & 0.99 & 1 & 0.97 & 0.76 & 0.68 & 0.65 & 0.38& 0.38& 0.38&0.24 &0.24 &0.24\\
Ours & 1 & 1 & 0.87 & 0.65 & 0.68 & 0.52 & 0.01& 0.01& 0& 0& 0.02& 0\\
\bottomrule
\end{tabular}}
\label{tab:guard-model}
\end{table*}


\paragraph{Comparison between the one-shot and many-shot black-box defenses.}\label{sec:selection}
We compare the performance of one-shot and many-shot black-box defenses in terms of DR on both PAIR and AutoDAN-Turbo attacks, as shown in Table~\ref{tab:shots}. Specifically, GPT-4.1 achieves a DR of 0.71 and 0.56 on PAIR and AutoDAN-T, respectively, under the many-shot setting, compared to 0.33 and 0.18 in the one-shot setting. Similarly, Gemini-2.5 performs better with a DR of 0.83 and 0.76 under many-shot defenses, compared to 0.27 and 0.22 in the one-shot case.

\begin{table}[htbp]
\centering
\caption{Comparison between the one-shot and many-shot black-box defenses.}
\label{tab:shots}
\resizebox{0.5\linewidth}{!}{
\begin{tabular}{lcccc}
\toprule
\multirow{2}{*}{\textbf{Models}} &
\multicolumn{2}{c}{\textbf{One-Shot}} &
\multicolumn{2}{c}{\textbf{Many-Shot}} \\
\cmidrule(lr){2-3} \cmidrule(lr){4-5}
& \textbf{PAIR} & \textbf{AutoDAN-T} & \textbf{PAIR} & \textbf{AutoDAN-T}  \\
\midrule
GPT-4.1 & 0.33 & 0.18& 0.71& 0.56\\
Gemini-2.5 & 0.27& 0.22 & 0.83& 0.76\\
\bottomrule
\end{tabular}
}
\end{table}

\paragraph{Progress of detection rate with a search budget of 300 iterations.}\label{sec:selection}

\begin{figure}[htbp]
    \centering
    \includegraphics[width=0.7\linewidth]{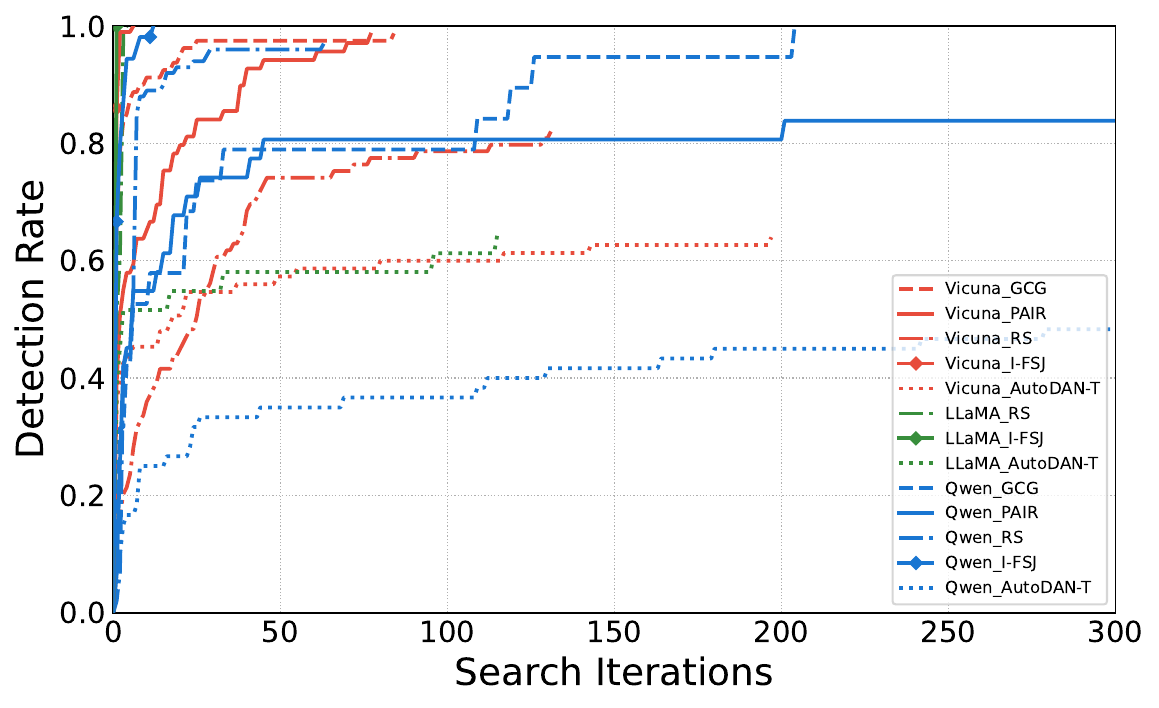}
    \caption{Detection rate evolution with respect to search iterations. The DR increases sharply in the initial iterations and continues to improve with a larger search budget.}
    \label{fig:efficient_random_300}
\vspace{-1.0em}    
\end{figure}

In Fig.~\ref{fig:efficient_random_300}, we show the evolution of the detection rate (DR) with respect to the number of search iterations. The process unfolds in two distinct stages: in the first stage ($\text{iter}<16$), the DR increases sharply, reflecting the high effectiveness of our guided search. Moreover, with a modest search budget (e.g., 50 iterations), the DR already reaches 90\%. With a larger search budget (e.g., 300 iterations), the DR continues to increase, as shown in Fig.~\ref{fig:efficient_random_300}.

%% file: arxiv.bib
@article{ziegler2019fine,
  title={Fine-tuning language models from human preferences},
  author={Ziegler, Daniel M and Stiennon, Nisan and Wu, Jeffrey and Brown, Tom B and Radford, Alec and Amodei, Dario and Christiano, Paul and Irving, Geoffrey},
  journal={arXiv preprint arXiv:1909.08593},
  year={2019}
}

@article{ouyang2022training,
  title={Training language models to follow instructions with human feedback},
  author={Ouyang, Long and Wu, Jeffrey and Jiang, Xu and Almeida, Diogo and Wainwright, Carroll and Mishkin, Pamela and Zhang, Chong and Agarwal, Sandhini and Slama, Katarina and Ray, Alex and others},
  journal={Advances in Neural Information Processing Systems},
  volume={35},
  pages={27730--27744},
  year={2022}
}

@article{bai2022constitutional,
  title={Constitutional ai: Harmlessness from ai feedback},
  author={Bai, Yuntao and Kadavath, Saurav and Kundu, Sandipan and Askell, Amanda and Kernion, Jackson and Jones, Andy and Chen, Anna and Goldie, Anna and Mirhoseini, Azalia and McKinnon, Cameron and others},
  journal={arXiv preprint arXiv:2212.08073},
  year={2022}
}

@inproceedings{korbak2023pretraining,
  title={Pretraining language models with human preferences},
  author={Korbak, Tomasz and Shi, Kejian and Chen, Angelica and Bhalerao, Rasika Vinayak and Buckley, Christopher and Phang, Jason and Bowman, Samuel R and Perez, Ethan},
  booktitle={International Conference on Machine Learning},
  pages={17506--17533},
  year={2023},
  organization={PMLR}
}

@article{wei2023jailbroken,
  title={Jailbroken: How does llm safety training fail?},
  author={Wei, Alexander and Haghtalab, Nika and Steinhardt, Jacob},
  journal={arXiv preprint arXiv:2307.02483},
  year={2023}
}

@article{zou2023universal,
  title={Universal and transferable adversarial attacks on aligned language models},
  author={Zou, Andy and Wang, Zifan and Kolter, J Zico and Fredrikson, Matt},
  journal={arXiv preprint arXiv:2307.15043},
  year={2023}
}

@misc{robey2024smoothllm,
      title={SmoothLLM: Defending Large Language Models Against Jailbreaking Attacks}, 
      author={Alexander Robey and Eric Wong and Hamed Hassani and George J. Pappas},
      year={2024},
      eprint={2310.03684},
      archivePrefix={arXiv},
      primaryClass={cs.LG},
      url={https://arxiv.org/abs/2310.03684}, 
}

@misc{jain2023baselinedefenses,
      title={Baseline Defenses for Adversarial Attacks Against Aligned Language Models}, 
      author={Neel Jain and Avi Schwarzschild and Yuxin Wen and Gowthami Somepalli and John Kirchenbauer and Ping-yeh Chiang and Micah Goldblum and Aniruddha Saha and Jonas Geiping and Tom Goldstein},
      year={2023},
      eprint={2309.00614},
      archivePrefix={arXiv},
      primaryClass={cs.LG},
      url={https://arxiv.org/abs/2309.00614}, 
}

@misc{alon2023Perplexity,
      title={Detecting Language Model Attacks with Perplexity}, 
      author={Gabriel Alon and Michael Kamfonas},
      year={2023},
      eprint={2308.14132},
      archivePrefix={arXiv},
      primaryClass={cs.CL},
      url={https://arxiv.org/abs/2308.14132}, 
}

@misc{kirchenbauer2024reliability,
      title={On the Reliability of Watermarks for Large Language Models}, 
      author={John Kirchenbauer and Jonas Geiping and Yuxin Wen and Manli Shu and Khalid Saifullah and Kezhi Kong and Kasun Fernando and Aniruddha Saha and Micah Goldblum and Tom Goldstein},
      year={2024},
      eprint={2306.04634},
      archivePrefix={arXiv},
      primaryClass={cs.LG},
      url={https://arxiv.org/abs/2306.04634}, 
}

@misc{provilkov2020bpedropout,
      title={BPE-Dropout: Simple and Effective Subword Regularization}, 
      author={Ivan Provilkov and Dmitrii Emelianenko and Elena Voita},
      year={2020},
      eprint={1910.13267},
      archivePrefix={arXiv},
      primaryClass={cs.CL},
      url={https://arxiv.org/abs/1910.13267}, 
}

@misc{mazeika2024harmbench,
      title={HarmBench: A Standardized Evaluation Framework for Automated Red Teaming and Robust Refusal}, 
      author={Mantas Mazeika and Long Phan and Xuwang Yin and Andy Zou and Zifan Wang and Norman Mu and Elham Sakhaee and Nathaniel Li and Steven Basart and Bo Li and David Forsyth and Dan Hendrycks},
      year={2024},
      eprint={2402.04249},
      archivePrefix={arXiv},
      primaryClass={cs.LG},
      url={https://arxiv.org/abs/2402.04249}, 
}

@misc{chao2024PAIR,
      title={Jailbreaking Black Box Large Language Models in Twenty Queries}, 
      author={Patrick Chao and Alexander Robey and Edgar Dobriban and Hamed Hassani and George J. Pappas and Eric Wong},
      year={2024},
      eprint={2310.08419},
      archivePrefix={arXiv},
      primaryClass={cs.LG},
      url={https://arxiv.org/abs/2310.08419}, 
}

@misc{kumar2025erasecheck,
      title={Certifying LLM Safety against Adversarial Prompting}, 
      author={Aounon Kumar and Chirag Agarwal and Suraj Srinivas and Aaron Jiaxun Li and Soheil Feizi and Himabindu Lakkaraju},
      year={2025},
      eprint={2309.02705},
      archivePrefix={arXiv},
      primaryClass={cs.CL},
      url={https://arxiv.org/abs/2309.02705}, 
}

@misc{zhou2023IF,
      title={Instruction-Following Evaluation for Large Language Models}, 
      author={Jeffrey Zhou and Tianjian Lu and Swaroop Mishra and Siddhartha Brahma and Sujoy Basu and Yi Luan and Denny Zhou and Le Hou},
      year={2023},
      eprint={2311.07911},
      archivePrefix={arXiv},
      primaryClass={cs.CL},
      url={https://arxiv.org/abs/2311.07911}, 
}

@misc{yang2021poisonedwordembeddings,
      title={Be Careful about Poisoned Word Embeddings: Exploring the Vulnerability of the Embedding Layers in NLP Models}, 
      author={Wenkai Yang and Lei Li and Zhiyuan Zhang and Xuancheng Ren and Xu Sun and Bin He},
      year={2021},
      eprint={2103.15543},
      archivePrefix={arXiv},
      primaryClass={cs.CL},
      url={https://arxiv.org/abs/2103.15543}, 
}

@misc{chao2024jailbreakbench,
      title={JailbreakBench: An Open Robustness Benchmark for Jailbreaking Large Language Models}, 
      author={Patrick Chao and Edoardo Debenedetti and Alexander Robey and Maksym Andriushchenko and Francesco Croce and Vikash Sehwag and Edgar Dobriban and Nicolas Flammarion and George J. Pappas and Florian Tramer and Hamed Hassani and Eric Wong},
      year={2024},
      eprint={2404.01318},
      archivePrefix={arXiv},
      primaryClass={cs.CR},
      url={https://arxiv.org/abs/2404.01318}, 
}

@misc{andriushchenko2025RS,
      title={Jailbreaking Leading Safety-Aligned LLMs with Simple Adaptive Attacks}, 
      author={Maksym Andriushchenko and Francesco Croce and Nicolas Flammarion},
      year={2025},
      eprint={2404.02151},
      archivePrefix={arXiv},
      primaryClass={cs.CR},
      url={https://arxiv.org/abs/2404.02151}, 
}

@misc{liu2025autodanturbo,
      title={AutoDAN-Turbo: A Lifelong Agent for Strategy Self-Exploration to Jailbreak LLMs}, 
      author={Xiaogeng Liu and Peiran Li and Edward Suh and Yevgeniy Vorobeychik and Zhuoqing Mao and Somesh Jha and Patrick McDaniel and Huan Sun and Bo Li and Chaowei Xiao},
      year={2025},
      eprint={2410.05295},
      archivePrefix={arXiv},
      primaryClass={cs.CR},
      url={https://arxiv.org/abs/2410.05295}, 
}

@misc{grattafiori2024llama3herdmodels,
      title={The Llama 3 Herd of Models}, 
      author={Llama Team},
      year={2024},
      eprint={2407.21783},
      archivePrefix={arXiv},
      primaryClass={cs.AI},
      url={https://arxiv.org/abs/2407.21783}, 
}

@misc{liu2025guardreasoner,
      title={GuardReasoner: Towards Reasoning-based LLM Safeguards}, 
      author={Yue Liu and Hongcheng Gao and Shengfang Zhai and Yufei He and Jun Xia and Zhengyu Hu and Yulin Chen and Xihong Yang and Jiaheng Zhang and Stan Z. Li and Hui Xiong and Bryan Hooi},
      year={2025},
      eprint={2501.18492},
      archivePrefix={arXiv},
      primaryClass={cs.CR},
      url={https://arxiv.org/abs/2501.18492}, 
}

@misc{hase2025smoothedembeddingsrobustlanguage,
      title={Smoothed Embeddings for Robust Language Models}, 
      author={Ryo Hase and Md Rafi Ur Rashid and Ashley Lewis and Jing Liu and Toshiaki Koike-Akino and Kieran Parsons and Ye Wang},
      year={2025},
      eprint={2501.16497},
      archivePrefix={arXiv},
      primaryClass={cs.LG},
      url={https://arxiv.org/abs/2501.16497}, 
}

@misc{dubois2025lengthcontrolledalpacaevalsimpleway,
      title={Length-Controlled AlpacaEval: A Simple Way to Debias Automatic Evaluators}, 
      author={Yann Dubois and Balázs Galambosi and Percy Liang and Tatsunori B. Hashimoto},
      year={2025},
      eprint={2404.04475},
      archivePrefix={arXiv},
      primaryClass={cs.LG},
      url={https://arxiv.org/abs/2404.04475}, 
}

@article{zheng2024improved,
  title={Improved few-shot jailbreaking can circumvent aligned language models and their defenses},
  author={Zheng, Xiaosen and Pang, Tianyu and Du, Chao and Liu, Qian and Jiang, Jing and Lin, Min},
  journal={Advances in Neural Information Processing Systems},
  volume={37},
  pages={32856--32887},
  year={2024}
}
